\documentclass[aps,twocolumn,pra,showpacs]{revtex4}
\usepackage{graphicx}
\usepackage{amssymb}
\usepackage{color}%
\usepackage{amsmath}%
\usepackage{dcolumn}
\usepackage{bm}
\usepackage[dvipdfm]{xcolor}
\usepackage{subfigure}

\usepackage{amsfonts}
\begin{document}
\title{ Cooperative scattering measurement of coherence in a spatially modulated Bose gas}
\author{Bo Lu}
\author{Thibault Vogt}
\email{thibault.vogt@pku.edu.cn}
\author{Xinxing Liu}
\author{Xu Xu}
\author{Xiaoji Zhou}
\email{xjzhou@pku.edu.cn}
\author{Xuzong Chen}
\email{xuzongchen@pku.edu.cn}
\affiliation{School of Electronics Engineering and Computer Science, Peking University, Beijing 100871, People's Republic of China}

\newcommand{\mb}[1]{\mathbf{#1}}

\begin{abstract}

Correlations of a Bose gas released from an optical lattice are measured using superradiant scattering.
Conditions are chosen so that after initial incident light pumping at the Bragg angle for diffraction, due to matter wave amplification and mode competition, superradiant scattering into the Bragg diffracted mode is preponderant. A temporal analysis of the superradiant scattering gain reveals periodical oscillations and damping due to the initial lack of coherence between lattice sites. Such damping is used for characterizing first order spatial correlations in our system with a precision of one lattice period.

\end{abstract}
\maketitle

The realization of Bose-Einstein condensation has rendered possible studies of completely coherent interactions between light and matter waves. The observation of superradiant scattering with a Bose-Einstein condensate (BEC) in an exemplary experiment by Ketterle et al. \cite{inouye} lead later on to the demonstration of amplification of matter maves \cite{inouye1999phase,kozuma1999phase,Inouye2}.
In such cooperative scattering processes, light scattering and subsequent atomic recoil are deeply modified and tend to be coherent and directional.
A method based on Raman superradiant scattering was recently used to analyse the coherence properties of a BEC \cite{sadler}. Exploiting the strong coherence dependence of the light and matter waves mixing occurring in superradiance \cite{kozuma1999phase,Inouye2}, it was able to spatially discriminate between normal and superfluid phases of the condensate.

No further studies have been carried out yet to estimate whether or not superradiance can be a good candidate for spatially characterizing coherence in more complicated
or correlated systems formed after loading Bose gases into optical lattices.
In principle, coherence properties can be accessed directly through interferences imaging. They can be accessed also via Bragg scattering \cite{ernst2009probing,Clement} or through time of flight imaging, though
indirectly since those techniques consist in measuring momentum distributions. Reconstruction of the full first order correlation function
using absorption imaging, not yet obtained, would actually rely on measurements of the momentum distribution as a function of time of flight \cite{zhang2009tomo}.
Studying the critical point of phase transitions and as a consequence fluctuations \cite{Bloch2000nat,greiner2,donner2007critical,sachdev2008quantum} require precise spatial resolution of correlation functions. In that trend, second order correlations have already been measured using Hanbury Brown-Twiss interferometry \cite{folling2005spatial}.
Accurate control of coherence properties will also be necessary for realizing quantum registers or quantum simulations with systems based on cold atoms trapped in optical
lattices \cite{zhou2010detecting}.
This control can consist in measuring phase correlations between lattice sites separated by $\Delta  \mb{x}$, i.e. the global coherence or spatially averaged first order
correlation function $\mathfrak{g} (\Delta \mb{x})=\overline{\psi^* \left(\mb{r}\right) \psi \left( \mb{r} + \Delta \mb{x}\right)}/\overline{|\psi \left(\mb{r}\right)|^2}$ \cite{Cohen}.
 This expression is true for a macroscopic wavefunction $\psi$, that may be approximated by $\psi(\mb{r})= \sqrt{N} \phi_0(\mb{r})\sum_n w(x-n d)$
 for the example of a sample of $N$ atoms in a 1D optical lattice at sufficiently high lattice depths, with $\phi_0(\mb{r})$ the macroscopic profile \cite{kramer2002macroscopic} and $w(x-nd)$ the Wannier function at site $x=nd$ along the optical lattice axis $x$, $d$
 standing for the lattice spacing.
For a slowly varying profile in comparison to $w$ and if $|\Delta \mb{x}|$ is an integral multiple of $d$ one finds $\mathfrak{g} (\Delta \mb{x})=\overline{\phi_0^* \left( \mb{r}\right) \phi_0 \left(\mb{r} + \Delta  \mb{x}\right)}$ because of the orthogonality of the
 Wannier functions.

\begin{figure}[hptb]
\begin{center}\includegraphics[width=7cm] {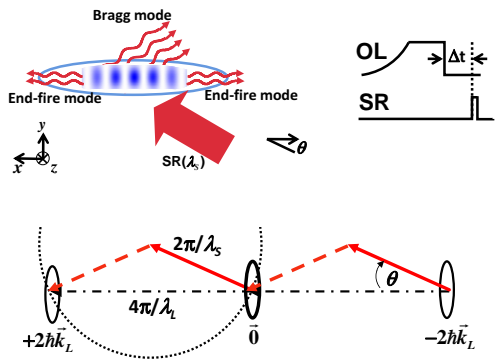}
\caption{(Color online) (top) Representation of the experiment. A cigar-shaped BEC is loaded along its long axis $x$ in an optical lattice (OL) formed by two counter-propagating beams ($\lambda_{L} \approx 852$ nm), which creates a matter grating. Following release of the optical lattice and a delay $\Delta t$, a $5\ \mu$s pump pulse (SR) is sent at $\theta=24^{\circ}$ with respect to the $x$ axis. This beam ($\lambda_S =780$~nm) can be superradiantly scattered into three modes, the Bragg mode symmetric of the incident beam with respect to the $x$ axis and two end-fire modes along the $x$ axis. (Bottom) Momentum representation of the Bose gas during the superradiance process.
Only the main recoil atomic momenta are presented on the picture. Absorption of one pump photon by the condensate at rest (momentum $\vec 0$) and subsequent emission at $24^{\circ}$ into the Bragg mode leads to amplification of the matter wave (MWA) with momentum $2 \hbar \vec k_L$.
The dotted circle stands for the energy conservation for atoms initially at rest during the Rayleigh scattering process.}%
\label{setup}%
\end{center}
\end{figure}

This article presents an alternative measurement of the global coherence function $\mathfrak{g}$ in a Bose gas loaded in a 1D optical lattice
 with a resolution of one lattice spacing.
 This study is based on superradiant scattering and employs the mechanism of amplification of matter waves (MWA), that can be efficiently driven in our system as was shown in a previous experiment on mode competition between superradiant scattering modes \cite{vogt}.
 Indeed, following release of a 1D optical lattice with period $\sim 0.5 \ \mu$m formed with retroreflected laser beams, three different atomic momenta are mostly populated, the Bose gas at rest $\phi_0(\mb{r})$ and two lightly populated waves $\alpha \phi_0( \mb{r} \pm  \mb{v} \Delta t)$ recoiling at $\mp \mb{v}$
 velocity, with $|\alpha|<<1$.
The angle and frequency of a pump pulse for superradiance can then be chosen at the Bragg angle for optical diffraction (see Fig. \ref{setup}), so that the condition of amplification of one of the wave is fulfilled
 \cite{Inouye2}. Since the MWA gain directly relates to atomic coherence between the atomic reservoir and the recoiling seed, it is closely related to $\mathfrak{g}(v \Delta t)$, where $\Delta t$ is the time before the pump pulse for MWA is applied.
Experimental analysis of this matter wave amplification process after release from the optical
lattice shows the MWA gain evolves actually periodically in time with a slow exponential damping due to lack of spatial coherence in the system. This result is used for characterizing coherence in the system while in the superfluide state.

Our experiment (see Fig.\ref{setup}) is performed with a cigar-shaped BEC of $N_t \sim 2\times 10^{5}$ $^{87}$Rb atoms in the $F=2, m_{F}=2$ state, with Thomas-Fermi length $70\ \mu m$ and width $7\ \mu$m \cite{YangFan,zhou}.
The BEC, obtained after evaporative cooling in a QUIC trap, is loaded along its long axis in an optical lattice formed with a retro-reflected laser beam ($\lambda_{L}=2\pi/k_L=852$ nm), focused on the BEC to a waist of $110\ \mu$m. A matter grating with period $d=\lambda_{L}/2$ forms because of the presence of the standing wave, whose optical lattice depth $V_0$ which we are to tune from 0 to $36\ E_{r}$, $E_{r}=\hbar^{2}k_{L}^2/2m$ being the recoil energy of a Rubidium atom. For an adiabatic loading to relatively high lattice depths, we use an exponential rising loading with $20$~ms constant and 40~ms rise time, and then we keep the optical lattice switched on for $50$ ms before sudden release of the combined optical and magnetic traps.
Following a short delay $\Delta t$, a light pulse, red-detuned from the D2 transition ($\lambda_{S} =780$ nm) by $\delta=$1.3~GHz, of duration $5\ \mu$s and polarized along the $z$ axis, is shined on the matter grating. The chosen angle $\theta \approx 24^{\circ}$ between the incident light direction and the direction of the long axis of the condensate is given by the Bragg's law $\cos\theta=\lambda_{s} / \lambda_{L}$ commonly used in X ray scattering. When this condition is satisfied, the pump pulse can be resonantly diffracted at $24^{\circ}$ in a direction which is the symmetric of the incident beam one with respect to the long axis of the BEC as shown in Fig. \ref{setup}.
Eventually, absorption imaging is performed after $30$ ms time of flight perpendicularly to the $x$ axis, from which we characterize the momentum distribution.
\begin{figure}[hptb]
\begin{center}
\includegraphics*[bb=88 265 296 575,angle=0,width=4.cm,height=3.8cm]{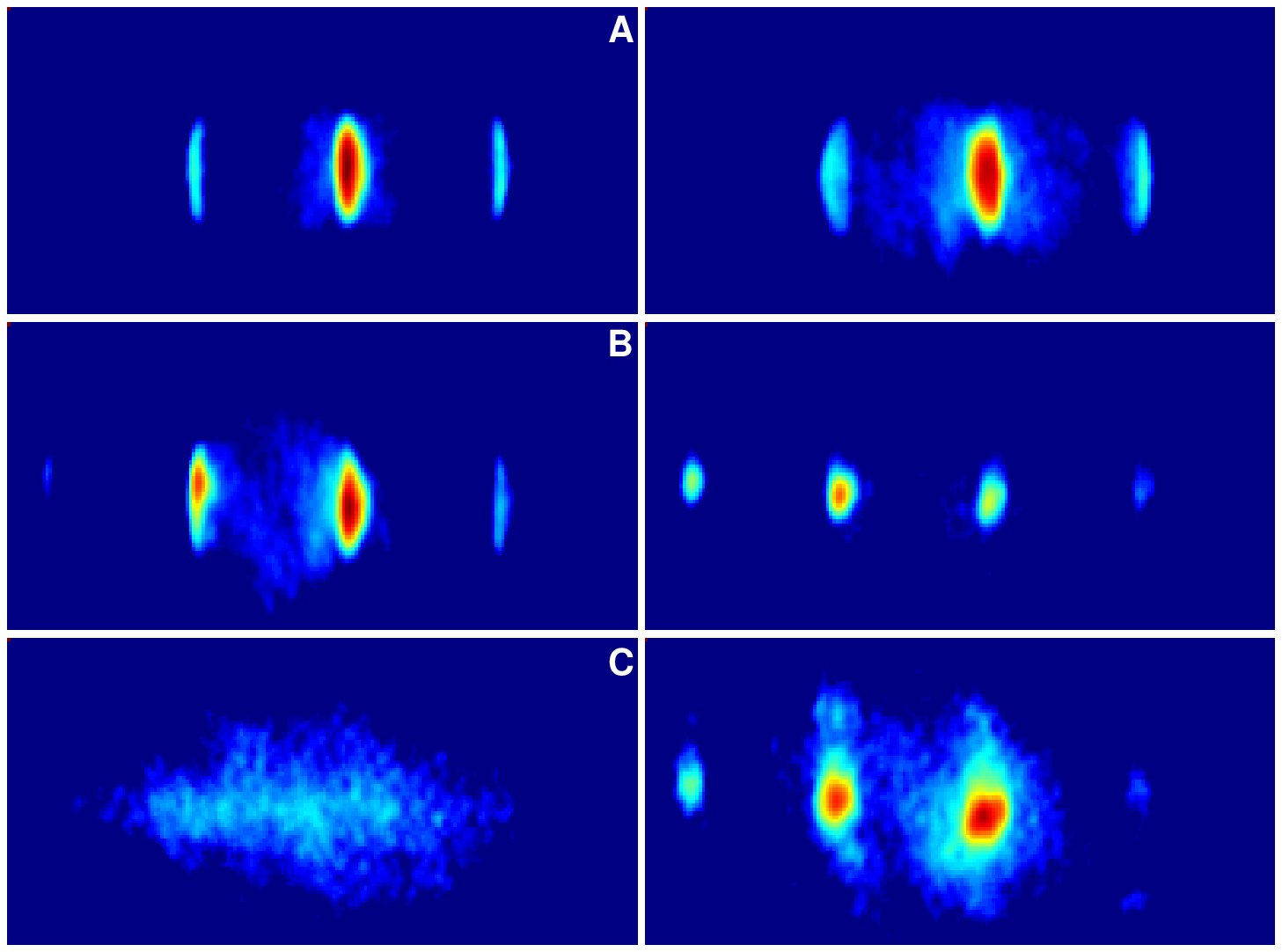}
\caption{(Color online) 30 ms time of flight absorption images. (A) Multiple matter-wave interference patterns
 after atoms were released from an optical lattice potential with a potential depth of $14.4\ E_{r}$.
 (B) MWA for an initial loading at $14.4\ E_{r}$ depth and $\Delta t=0$.
 (C) MWA at $36\ E_{r}$ lattice depth and $\Delta t=0$. For sake of visibility, the upper scaling value in Fig. (C) is divided by 2.}%
\label{tof}%
\end{center}
\end{figure}
Different absorption images are shown in Fig. \ref{tof}, for experiments performed with $\Delta t=0$.
In (A) is shown the multiple matter-wave interference patterns after atoms were released from an optical lattice potential with $V_0=14.4\ E_{r}$.
In (B), when adding the pump pulse after release from the optical lattice, a strong enhancement of the $+2\hbar  \mb{k}_{L}$ mode is observed because of MWA, while the  0 and $-2\hbar  \mb{k}_{L}$ orders are depleted. In (C), MWA for $V_0=36\ E_{r}$ appears still efficient, though the interference pattern is blurred because a complete loss of coherence occurs as the lattice depth is increased \cite{orzel2001squeezed}.
In principle, usual Rayleigh scattering superradiance (SR0) with light superradiant scattering into end-fire modes along the $x$ axis could also be observed. Nevertheless, because of the large seeding of the $+2\hbar  \mb{k}_{L}$ mode, SR0 is completely negligible as a result of mode competition for lattice depths we consider in this article, i.e. $V_0>5 E_{r}$ \cite{vogt}.

For further explaining how to characterize coherence, interpretation of the data is carried out using the quantum model of references \cite{Pu,vogt}  modified to take into account coherence.
In the superfluid regime, the one-particle wavefunction of the system at time $t=0$ immediately after lattice switch off is $\psi( \mb{r})=\sum_n w(x-nd)\psi_L(nd,y,z) \approx \psi_L( \mb{r}) \left(\tilde{w}(0) + \tilde{w}(-2  \mb{k}_L) e^{2ik_L x} + \tilde{w}( 2  \mb{k}_L) e^{-2ik_L x} \right)$, where $\psi_L( \mb{r})$ is the Fourier transform of the quasi-momentum distribution inside the optical lattice and $\tilde{w}$ designates the Fourier transform of the Wannier function $w(x)$.
The matter wave field is limited to the first scattering orders, thus expressed as $\hat{\Psi}( \mb{r},t)=\sum_{ \mb{q}} \phi_{ \mb{q}}( \mb{r},t) e^{i  \mb{q}  \mb{r}} \hat{c}_{ \mb{q}}(t)$ summed over $\hat{c}_0$, $\hat{c}_{2 \mb{k}_L}$ and $\hat{c}_{-2 \mb{k}_L}$, the annihilation mode operators of atoms with corresponding momenta and profiles $\phi_{ \mb{q}}( \mb{r},t)$. Considering such profiles as approximately given by the free evolution of  the initial wavefunction $\psi( \mb{r})$ during time $t$,  we have $\phi_{\pm2 \mb{k}_L}( \mb{r},t)= \phi_{0}( \mb{r}\mp  \mb{v} t,t)$ with $\phi_{0}( \mb{r},0)= \psi_L( \mb{r})$. Here, $ \mb{v}=\hbar 2  \mb{k}_L /m$ is the velocity at which the atoms recoil in mode $ 2\mb{ k}_L$, related to the frequency of recoil $\omega=4 \hbar k_L^2/2m$. The orthogonality relation $[\hat{c}_{ \mb{q}},\hat{c}_{ \mb{q'}}^\dagger]\approx \delta_{ \mb{q},  \mb{q'}}$ is verified when the quasi-momentum spread $2\Delta k$ is smaller than $2 k_L$, $\Delta  k< k_L$, which is valid if assuming only the lowest band is populated inside the optical lattice. The time evolution of the mode operators reads:
\begin{align}
\dot{\hat{c}}_{\pm 2  \mb{k}_L}= -i \omega \hat{c}_{\pm 2  \mb{k}_L} \pm \frac{N_0}{2}  \left(G_1\hat{c}_{\pm 2 \mb{k}_L}+G_2\hat{c}^{\dagger}_{\mp 2  \mb{k}_L}\right)
\label{rate}
\end{align}
In this expression, we have replaced $\hat{c}_0^{\dagger}\hat{c}_0$ by $N_0$ the number of atoms in the condensate at rest, that depends on time because of depletion during the superradiant process. As the number of atoms $N_{\pm 2 \mb{k}_L}$ of atoms in $\pm 2 \mb{k}_L$ modes are large, one replaces also the operators $\hat{c}_{\pm 2\mb{k_L}}$ by c-numbers of the form $c_{\pm 2\mb{k}_L}=\sqrt{N_{\pm 2 \mb{k_L}}}e^{i \varphi (t)}$ with initial conditions $c_{\pm 2\mb{k}_L}\left( 0 \right) = \sqrt{N_{\pm 2 \mb{k}_L}}$ and phase given by $\varphi (t)= -\omega t$ during the free expansion time. $G_1(t) = \int d \mb{k} |\beta(\mb{k})|^2 |\rho(\mb{k},t)|^2\delta[|\mb{k}|-|\mb{k}_S|]$ and $G_2(t)=\int d  \mb{k} |\beta( \mb{k})|^2 |\rho(\mb{k},t)|^2e^{-i( \mb{k} -\mb{k}_S+2\mb{k}_L)\cdot \mb{v} t}\delta[|\mb{k}|-|\mb{k}_S|]$
 are the gains for MWA into mode $-2 \mb{k}_L$.
 They depend on the overlap integral $\rho(\mb{k},t) = \int d \mb{r} \phi^*_0(\mb{r}-\mb{v} t,t)\phi_0(\mb{r},t)\exp[-i(\mb{k}-\mb{k}_S+2 \mb{k}_L)\cdot \mb{r}]$, centered at $\mb{k}_S-2 \mb{k}_L$, and on $\beta(\mb{k})=\frac{|\Omega_0 \mu|}{2|\delta|}(\frac{c k_S}{2\hbar\varepsilon_{0}(2\pi)^{3}})^{1/2}|\mb{k}\times \mb{z}|$
the atomic coupling coefficient between pump and scattered light, $\mu$ and $\Omega_0$ referring, respectively, to the dipole moment of the transition involved and its Rabi frequency taken as a step-like pump pulse starting at time $t=\Delta t$.

The possibility of retrieving the function $\mathfrak{g}$ comes from the MWA dependence on the delay time $\Delta t$. Indeed, one can show the identity
$\rho(\mb{k}_S-2\mb{k}_L, \Delta t)=\int d \mb{r} \phi^*_0(\mb{r}-\mb{v}\Delta t,\Delta t)\phi_0(\mb{r},\Delta t)=\mathfrak{g}(v\Delta t)$
which shows the gain for an exact transfer of $2\hbar \mb{k}_L$ momentum, $ |\beta (\mb{k}_S)\rho(\mb{k}_S-2\mb{k}_L)|^2$, is not affected by the expansion of the profile $\phi_0(\mb{r},\Delta t)$ but only by the coherence $\mathfrak{g}(v\Delta t)$.
If one assumes $\phi_0(\mb{r},0)=f_0(y,z)g_0(x)$, where $f_0$ is a gaussian function independent on $x$ or to the least varying very slowly with $x$ and $g_0$ depends only on $x$, the MWA gain of $+2\hbar \mb{k}_{L}$ mode, corresponding to an integration over the momentum distributions of the gains $ |\beta (\mb{k}_S) \rho(\mb{k}, \Delta t)|^2$ with $k=k_S$, takes the form
\begin{equation}
 G_1 (\Delta t)\approx A |\beta (\mb{k}_S)|^2 \int {\rm d} x g^*_1(x-v \Delta t,x)g_{1m}(x-v \Delta t,x)\notag
\end{equation}
 where $A$ is a constant and $g_{1m}(x-v \Delta t,x)=\int {\rm d} u g_1(u+x-v \Delta t,u+x)e^{\frac{-u^2 \tan^2\theta_0}{2\sigma^2}}$, $g_1$ being given by $g_1(x-v\Delta t,x)=g_0^*(x-v\Delta t)g_0(x)$ and $\sqrt{2} \sigma$ being the radius of the Bose gas inside the optical lattice, well approximated by the Thomas-Fermi radius in the superfluide state. The Gaussian window function $e^{\frac{-u^2 \tan^2\theta_0}{2\sigma^2}}$ that stems from the $\mb{k}_x$ dispersion of the superradiantly emitted light, varies by less than 10$\%$ over 2$\sigma$ (our case $12$ lattice spacings $d$). Thus $g_{1m}(x-v \Delta t,x)$ provides a local measure of coherence at position $x$ and taken in $v\Delta t$ when this window is larger than the local coherence length $L$.
 For the case of an ergodic system, the local coherence function is equivalent to the global coherence, with consequence
 $G_1(\Delta t)\propto|\mathfrak{g}(v\Delta t)|^2$. In reality, this approximation is not completely exact but even in case of sharp spatial variations of the state of the system with spatial variations of the local coherence, the function $G_1$ still provides a good qualitative measure of the global coherence.
The variation of the amplitude of the MWA gain with time $\Delta t$ is then entirely contained in the $G_1$ function, from which coherence between positions separated by $n$ lattice sites can be obtained when $\Delta t= n T_0$ with $T_0= d/v=\pi/\omega \approx 40\ \mu$s the time for a recoil over a distance equal to one lattice site $d$. $G_2$ can be derived in the same manner, yielding $G_2 (\Delta t)=A |\beta (\mb{k}_S)|^2 \int {\rm d} x g^*_1(x-v \Delta t,x)g_{1m}(x,x+v \Delta t)$. We can assume here $g_{1m}(x,x+v \Delta t)\approx g_{1m}(x-v \Delta t,x)$ since the distances $v \Delta t$ that give a non zero gain are anyways smaller than the coherence length, so that we define $G=G_1 \approx G_2$. Our model does not consider Doppler dephasing during the very short superradiance pumping, as it appears in Eq. (\ref{rate}) with a unique recoiling frequency $\omega$. Doppler dephasing occurring before pumping can be discarded too while deriving the expression for $G_1$ as we are not interested in precision better than $d$ in $g_0(x)$. As can be shown rigorously, this is valid if the spatial spread induced by the velocity dispersion verifies $v_{\Delta k} \Delta t<d$, which for the example of an exponential damping of coherence with $\Delta k\sim 2/L$ only requires $v\Delta t <\pi L$.

\begin{figure}
\begin{center}
    \includegraphics[
    trim=0.317422in 0.347189in 0.113456in 0.125724in,
    width=7.5cm]{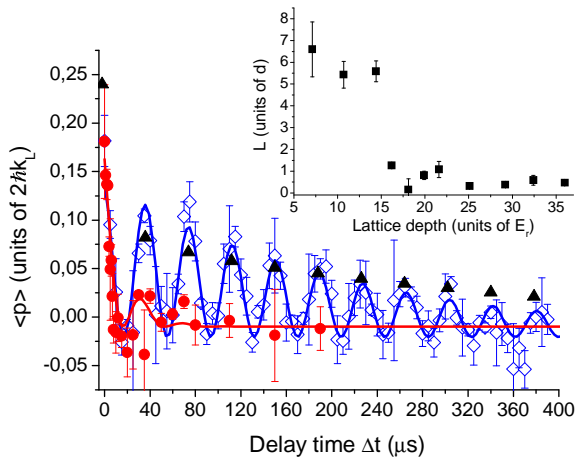}
     \end{center}
\caption{(Color online) (bottom) Average momentum $\langle p\rangle$ as a function of delay time $\Delta t$, calculated from time of flight images. The optical lattice depths are $14.4\ E_{r}$ (empty blue diamonds) and $36\ E_{r}$ (filled red circles). Error bars are obtained after averaging over several experimental data. Solid lines are empirical fits (see text). The black triangle points are a theoretical calculation of the global coherence function (rescaled amplitude). (Inset, top) Dependence of the correlation length $L$ versus the optical lattice depth obtained from the empirical fits of the data. Error bars indicate the 70~$\%$ confidence bound result from these temporal fits.}%
\label{vsdelay_400mVand1100mV}
\end{figure}

To verify this coherence dependence, we impinged a short MWA light pulse of $5\ \mu$s on the matter wave grating after different delays $\Delta t$ ranging from $0$ to $400\ \mu$s and for different lattice depths.
From each time of flight image, we compute the average momentum along $x$, $\langle p \rangle=\int n(\mb{p}_x) \mb{p}_x d \mb{p}_x \approx 2\hbar k_{L} \left( N_{2\mb{k}_L}-N_{-2\mb{k}_L}\right)/N_t$. Such definition characterizes the average momentum change due to MWA even at high lattice depth, when different initial momenta are populated (cf Fig. \ref{tof} (C)).
In Fig.\ref{vsdelay_400mVand1100mV}, regular oscillations of $\langle p \rangle$ are observed as a function of delay time for a 14.4~$E_{r}$ lattice depth. Damping appears on a timescale of the order of 200$\ \mu$s.
To retrieve the damping time $\tau$ and the oscillation period $T$, the data are fitted according to the heuristic formula $A+B \exp(-\Delta t/\tau) \cos(2\pi \Delta t/T+\delta \phi_0)$, the initial phase $\delta \phi_0$ being attributed to the finite duration of the pump pulse. We find $T \approx T_0$ for every lattice depth.
Such oscillations of the gain are directly linked to light diffraction by the matter grating, which is maximum only when the recoiling modes $-2\hbar \mb{k}_L$ and $2\hbar \mb{k}_L$ are in phase.
According to Eq. (\ref{rate}), for small depletion of the condensate at rest and short pump pulses ($T_p$ duration), the average momentum is directly proportional to $G (\Delta t)$ and effectively oscillates with period $T_0$:
\begin{equation}
\langle p \rangle=4\hbar k_{L}  T_p \ G(\Delta t) \left(1+\cos\left(2 \omega \Delta t \right)\right)
\label{momentum}
\end{equation}
We carried out a systematic measurement of the damping time of oscillations versus the optical lattice depth (cf inset in Fig.\ref{vsdelay_400mVand1100mV}).
Of the order of $200\ \mu$s in the superfluid state, a fast drop of the damping time is observed between $15$ and $20\ E_{r}$, after which it is nearly zero.
If the lattice depth is increased to 36~$E_{r}$ for example (see Fig.\ref{vsdelay_400mVand1100mV}), almost no oscillation is observed, the damping time is indeed shorter than one period.
From Eq. (\ref{momentum}), we know this damping is directly related to the coherence of the system.
We can infer a coherence length defined as $L=v\tau$ to be larger than $5\ d$ for $V_0=14.4\ E_{r}$. If our interpretation, which is valid only in the superfluid state, could be extended to higher lattice depths, that would mean $L$ is smaller than $d$ for $V_0 > 17\ E_{r}$.

For comparing to theory, a coherence function proportional to $|\mathfrak{g}(v\Delta t)|^2$ is simulated for a 3D BEC of non interacting bosons loaded in a 1D optical lattice in the superfluid phase, combining the models developed in ref. \cite{Lin2008} and \cite{rey2005ultracold} for the calculation of the system wavefunction.
Initial finite temperature of our BEC, $T_b\sim120$ nK, is taken into account, so that limitation of the coherence comes from incoherent population of the different states, the magnetic trap levels being also considered for more precision. The computed coherence function is displayed in Fig. \ref{vsdelay_400mVand1100mV} for $V_0=14.4\ E_r$.
A relatively good agreement with the experimental results is observed in the superfluid phase. Smaller coherence at long distances is measured experimentally, probably due to the presence of interactions which should be considered in our simulation. Our theory cannot account for the fast drop observed at high lattice depths either. This drop can be interpreted as a complete loss of coherence when the lattice depth is increased above a rather sharp threshold between superfluid and atomic squeezed state \cite{orzel2001squeezed}. Large radial confinement at high lattice depths may also induce long-time scale redistribution of the atoms and consequently non adiabatic loading, heating and loss of coherence \cite{mckagan2006mean}.
Further theoretical analysis will be necessary to determine the ultimate precision of our measurement of the correlation function and whether or not it is still valid above the observed threshold.

In conclusion, MWA was used to characterize phase correlations of a Bose gas initially loaded in an optical lattice.
A quantum model was developed to account for the main features observed in the experiment and confirmed our ability to measure phase correlations with an accuracy of one lattice spacing in the superfluid phase.
This method is relatively easy to implement. It may be used to characterize quantum phase transitions
in well controlled systems \cite{greiner2}, for example when magnetic trap frequencies are relaxed to get less dense condensate and with 3D optical lattices. Superradiant
light scattering rather than absorption imaging could be used as a direct way for measuring MWA gain.

We are grateful to Cheng Chin and Wolfgang Ketterle for their comments and Wei Zhang for his help on simulations. This work is partially supported
by the state Key Development Program for Basic Research
of China (Grants No. 2005CB724503, No. 2006CB921402,
and No. 2006CB921401) and by NSFC (Grants No. 10874008 and
No. 10934010).



\end{document}